\documentstyle[12pt]{article}

\catcode`\@=11
\long\def\@makefntext#1{
\protect\noindent \hbox to 3.2pt {\hskip-.9pt
$^{{\ninerm\@thefnmark}}$\hfil}#1\hfill}		

\def\@makefnmark{\hbox to 0pt{$^{\@thefnmark}$\hss}}  

\def\ps@myheadings{\let\@mkboth\@gobbletwo
\def\@oddhead{\hbox{}
\rightmark\hfil\ninerm\thepage}
\def\@oddfoot{}\def\@evenhead{\ninerm\thepage\hfil
\leftmark\hbox{}}\def\@evenfoot{}
\def\sectionmark##1{}\def\subsectionmark##1{}}

\renewcommand{\thefootnote}{\fnsymbol{footnote}}

\newcounter{sectionc}\newcounter{subsectionc}\newcounter{subsubsectionc}
\renewcommand{\section}[1] {\vspace*{0.6cm}\addtocounter{sectionc}{1}
\setcounter{subsectionc}{0}\setcounter{subsubsectionc}{0}\noindent
	{\normalsize\bf\thesectionc. #1}\par\vspace*{0.4cm}}
\renewcommand{\subsection}[1] {\vspace*{0.6cm}\addtocounter{subsectionc}{1}
	\setcounter{subsubsectionc}{0}\noindent
	{\normalsize\it\thesectionc.\thesubsectionc. #1}\par\vspace*{0.4cm}}
\renewcommand{\subsubsection}[1]
{\vspace*{0.6cm}\addtocounter{subsubsectionc}{1}
	\noindent {\normalsize\rm\thesectionc.\thesubsectionc.\thesubsubsectionc.
	#1}\par\vspace*{0.4cm}}

\newcounter{appendixc}
\newcounter{subappendixc}[appendixc]
\newcounter{subsubappendixc}[subappendixc]

\renewcommand{\appendix}[1] {\vspace*{0.6cm}
        \refstepcounter{appendixc}
        \setcounter{figure}{0}
        \setcounter{table}{0}
        \setcounter{equation}{0}
        \renewcommand{\thefigure}{\Alph{appendixc}.\arabic{figure}}
        \renewcommand{\thetable}{\Alph{appendixc}.\arabic{table}}
        \renewcommand{\theappendixc}{\Alph{appendixc}}
        \renewcommand{\theequation}{\Alph{appendixc}.\arabic{equation}}
        \noindent{\bf Appendix \theappendixc #1}\par\vspace*{0.4cm}}

\def\abstracts#1{{
	\centering{\begin{minipage}{12.2truecm}
\footnotesize\baselineskip=12pt\noindent
	\centerline{\footnotesize ABSTRACT}\vspace*{0.3cm}
	\parindent=0pt #1
	\end{minipage}}\par}}

\renewenvironment{thebibliography}[1]
	{\begin{list}{\arabic{enumi}.}
	{\usecounter{enumi}\setlength{\parsep}{0pt}
\setlength{\leftmargin 1.25cm}{\rightmargin 0pt}
	 \setlength{\itemsep}{0pt} \settowidth
	{\labelwidth}{#1.}\sloppy}}{\end{list}}

\newcounter{itemlistc}
\newcounter{romanlistc}
\newcounter{alphlistc}
\newcounter{arabiclistc}

\newcommand{\fcaption}[1]{
        \refstepcounter{figure}
        \setbox\@tempboxa = \hbox{\footnotesize Fig.~\thefigure. #1}
        \ifdim \wd\@tempboxa > 6in
           {\begin{center}
        \parbox{6in}{\footnotesize\baselineskip=12pt Fig.~\thefigure. #1}
            \end{center}}
        \else
             {\begin{center}
             {\footnotesize Fig.~\thefigure. #1}
              \end{center}}
        \fi}

\newcommand{\tcaption}[1]{
        \refstepcounter{table}
        \setbox\@tempboxa = \hbox{\footnotesize Table~\thetable. #1}
        \ifdim \wd\@tempboxa > 6in
           {\begin{center}
        \parbox{6in}{\footnotesize\baselineskip=12pt Table~\thetable. #1}
            \end{center}}
        \else
             {\begin{center}
             {\footnotesize Table~\thetable. #1}
              \end{center}}
        \fi}

\def\@citex[#1]#2{\if@filesw\immediate\write\@auxout
	{\string\citation{#2}}\fi
\def\@citea{}\@cite{\@for\@citeb:=#2\do
	{\@citea\def\@citea{,}\@ifundefined
	{b@\@citeb}{{\bf ?}\@warning
	{Citation `\@citeb' on page \thepage \space undefined}}
	{\csname b@\@citeb\endcsname}}}{#1}}

\newif\if@cghi
\def\cite{\@cghitrue\@ifnextchar [{\@tempswatrue
	\@citex}{\@tempswafalse\@citex[]}}
\def\citelow{\@cghifalse\@ifnextchar [{\@tempswatrue
	\@citex}{\@tempswafalse\@citex[]}}
\def\@cite#1#2{{$\null^{#1}$\if@tempswa\typeout
	{IJCGA warning: optional citation argument
	ignored: `#2'} \fi}}

 1
 1
 1

\font\ninerm=cmr9

\newcommand{\newc}{\newcommand}

\newc{\vs}{{\it vs.}}
\newc{\msusy}{M_{\rm SUSY}}
\newc{\wgratio}{x}
\newc{\ssqthw}{\sin^2\theta_W}
\newc{\Rb}{R_b}
\newc{\mone}{M_1}
\newc{\mtwo}{M_2}
\newc{\alphasmz}{\alpha_s(\mz)}
\newc{\alphasmzmin}{\alphas^{\rm min}(\mz)}

\newc{\slepton}{{\tilde{l}}}	\newc{\mslepton}{m_\slepton}
\newc{\supq}{{\tilde{u}}}
\newc{\sdown}{{\tilde{d}}}
\newc{\selectron}{{\tilde{e}}}
\newc{\sneutrino}{{\tilde{\nu}}}
\newc{\squark}{{\tilde{q}}}	\newc{\msquark}{m_\squark}
\newc{\higgsino}{{\tilde{H}}}

\newc{\mtop}{\mt}
\newc{\btau}{$b$--$\tau$}
\newc{\bino}{\widetilde B}
\newc{\wino}{\widetilde W}
\newc{\beq}{\begin{equation}}
\newc{\eeq}{\end{equation}}
\newc{\bea}{\begin{eqnarray}}
\newc{\eea}{\end{eqnarray}}
\newc{\onehalf}{\frac{1}{2}}
\newc{\gsim}{\lower.7ex\hbox{$\;\stackrel{\textstyle>}{\sim}\;$}}
\newc{\lsim}{\lower.7ex\hbox{$\;\stackrel{\textstyle<}{\sim}\;$}}
\newc{\alphas}{\alpha_s}
\newc{\tanb}{\tan\beta}
\newc{\mz}{m_Z}		\newc{\mw}{m_W}
\newc{\mhalf}{m_{1/2}}
\newc{\mzero}{m_0}
\newc{\muzero}{\mu_0}
\newc{\sgnmu}{{\rm sgn}\,\muzero}
\newc{\azero}{A_0}
\newc{\Atop}{A_t}
\newc{\bzero}{B_0}
\newc{\mt}{m_t}
\newc{\mb}{m_b}
\newc{\mtau}{m_\tau}
\newc{\mq}{m_q}
\newc{\htop}{h_t}
\newc{\hbot}{h_b}
\newc{\htau}{h_\tau}
\newc{\mtpole}{M_t}
\newc{\mbpole}{M_b}
\newc{\mqpole}{M_q}
\newc{\mgut}{M_X}
\newc{\mx}{\mgut}
\newc{\alphax}{\alpha_X}
\newc{\ie}{{\it i.e.}}
\newc{\etal}{{\it et al.}}
\newc{\eg}{{\it e.g.}}
\newc{\etc}{{\it etc.}}
\newc{\hh}{{h^0}}
\newc{\mhh}{m_\hh}
\newc{\hH}{{H^0}}
\newc{\mhH}{m_\hH}
\newc{\hA}{{A^0}}
\newc{\mhA}{m_\hA}
\newc{\hpm}{{H^\pm}}
\newc{\mhpm}{m_\hpm}
\newc{\stp}{{\widetilde t}}
\newc{\stopl}{{\stp_L}}		\newc{\mstopl}{m_{\stopl}}
\newc{\stopr}{{\stp_R}}		\newc{\mstopr}{m_{\stopr}}
\newc{\stau}{{\widetilde\tau}}
\newc{\gluino}{{\widetilde g}}	\newc{\mgluino}{m_{\gluino}}
\newc{\MS}{{\rm\overline{MS}}}
\newc{\DR}{{\rm\overline{DR}}}
\newc{\ev}{{\rm\,eV}}
\newc{\mev}{{\rm\,MeV}}
\newc{\gev}{{\rm\,GeV}}
\newc{\tev}{{\rm\,TeV}}
\newc{\bsg}{BR$(b\to s\gamma)$}
\newc{\abundchi}{\Omega_\chi h_0^2}
\newc{\mchi}{m_\chi}
\newc{\mcharone}{m_{\charone}}
	\newc{\charone}{\chi_1^\pm}
\newc{\mneutone}{m_{\neutone}}	\newc{\neutone}{\chi^0_1}
\newc{\mneuttwo}{m_{\neuttwo}}	\newc{\neuttwo}{\chi^0_2}
\def\NPB#1#2#3{{\it Nucl. Phys.} {\bf B#1} (19#2) #3}
\def\PLB#1#2#3{{\it Phys. Lett.} {\bf B#1} (19#2) #3}

\def\PRD#1#2#3{{\it Phys. Rev.} {\bf D#1} (19#2) #3}

\def\PRT#1#2#3{{\it Phys. Rep.}  {\bf#1} (19#2) #3}

\begin{document}

\newcommand{\st}{\scriptstyle}
\newcommand{\sst}{\scriptscriptstyle}
\newcommand{\mco}{\multicolumn}
\newcommand{\epp}{\epsilon^{\prime}}
\newcommand{\vep}{\varepsilon}
\newcommand{\ra}{\rightarrow}
\newcommand{\ppg}{\pi^+\pi^-\gamma}
\newcommand{\vp}{{\bf p}}
\newcommand{\ko}{K^0}
\newcommand{\kb}{\bar{K^0}}
\newcommand{\al}{\alpha}
\newcommand{\ab}{\bar{\alpha}}
\def\be{\begin{equation}}
\def\ee{\end{equation}}
\def\bea{\begin{eqnarray}}
\def\eea{\end{eqnarray}}
\def\CPbar{\hbox{{\rm CP}\hskip-1.80em{/}}}

\begin{center} \Large
{\bf Theoretical Physics Institute}\\
{\bf University of Minnesota}
\end{center}
\begin{flushright}
TPI-MINN-95/06-T\\
UMN-TH-1335-95\\
hep-ph/9503487\\
March 1995
\end{flushright}
\vspace{.3cm}

\centerline{\normalsize\bf }
\baselineskip=22pt
\centerline{\normalsize\bf SUPERSYMMETRIC PHENOMENOLOGY}
\baselineskip=16pt
\centerline{\normalsize\bf IN THE LIGHT OF GRAND UNIFICATION}

\centerline{\footnotesize LESZEK ROSZKOWSKI\footnote{Plenary talk at the
{\it Conference on Beyond the Standard Model IV}, Lake
Tahoe, CA, Dec. 13-18, 1994,
to appear in the Proceedings.
}
}
\baselineskip=13pt
\centerline{\footnotesize\it Department of Physics, University of Minnesota}
\baselineskip=12pt
\centerline{\footnotesize\it Minneapolis, MN 55455}
\centerline{\footnotesize E-mail: leszek@mnhepw.hep.umn.edu}

\vspace*{0.9cm}
\abstracts{
I review some aspects of supersymmetric grand unification and
emphasize a recent development in the area of gauge coupling unification.
}

\normalsize\baselineskip=15pt
\setcounter{footnote}{0}
\renewcommand{\thefootnote}{\alph{footnote}}

\section{Introduction}
During the last
few years supersymmetry (SUSY) has gained the status of the most
likely
candidate for
physics beyond the Standard Model (SM). The very fact that so many
talks at this meeting are devoted to various aspects of
supersymmetric physics, and so
few to its fading alternatives, is the best illustration of the
current situation. This ``SUSY fervor'' might be considered
unjustified in light of the fact that, despite many experimental
efforts, no signal of SUSY has been detected. Still, as I will demonstrate
below, there are good reasons to like SUSY. Introduced to particle
physics over a decade ago in order to make sense of grand unified theories
(GUT's), SUSY was subsequently shown to possess many other remarkable
features, like allowing for unifying
gravity
with other interactions, or providing an excellent dark matter candidate, to
name just a few.
SUSY GUT's also correctly predicted the value of
$\ssqthw$, where $\theta_{\rm W}$ is the weak
angle.

The development of the last few years is well known. Around 1989 LEP and
SLC provided precise measurements of the gauge couplings of the SM.
Then it became clear\cite{earlyunif} that, in the $\MS$ scheme, the three
(running) gauge coupling do not unify anywhere in the SM but they
nicely do so at an expected scale $\mgut\sim10^{16}\gev$
in the Minimal Supersymmetric Standard Model
(MSSM).\cite{minisugra} This was very encouraging news for both
SUSY and GUT's (or string theory) since otherwise
each of them lacks real theoretical strength and motivation without
the other.

Further studies expanded the guiding idea of unification into several
related directions. Basically, what was studied, by many groups and at various
levels of sophistication, was:
\begin{itemize}
\item gauge coupling unification;
\item Yukawa coupling unification;
\item mass unification.
\end{itemize}

I will discuss these topics in turn. Of course, each of them has grown
into a rich and impressive subject of its own and in preparing this talk
I had to make some hard choices. Thus instead of attempting to even
briefly mention every result which bears at least some significance, I will
rather sketch the overall picture and select a few
most, in my biased opinion, characteristic results.
Initially, my plan was to spend little time on the
first two topics, since they had been studied somewhat
earlier, had already been
presented at various meetings, and, it seemed to me,
relatively little work had been done on them recently. As it came out, some
new and interesting development took place during the last
few months in the area of gauge coupling
unification. (In fact, some progress has taken place after the
conference -- I will update my talk accordingly.)
But first I will
briefly set the stage on which this unification game is usually played.

\section{Framework}

The simplest and most popular framework for studying SUSY is the MSSM
(with $R$-parity assumed).
Since, viewed as a mere phenomenological extension of the
SM, the MSSM contains many unknown parameters, usually some
GUT-physics connection is adopted. Most commonly one assumes
that soft terms needed to break SUSY are generated by coupling the
MSSM to minimal $N=1$ supergravity, and additionally a particularly
simple (delta-like) form of the kinetic term for the gauge superfields
is chosen, and a simple
unification group is also assumed.\cite{minisugra}
This leads to several
mass relations. For example, the gauginos -- the bino of
$U(1)_Y$, the winos of $SU(2)_L$, and the gluinos of $SU(3)_c$ have
equal masses at the GUT (actually, Planck) scale:
$\mone=\mtwo=\mgluino=\mhalf$. This leads, due to renormalization
effects, to the following well-known relations at the electroweak scale:

\bea
\mone&=& {5\over3}\tan^2\theta_{\rm W}\,\mtwo\simeq0.5 \mtwo,
\label{monemtwo:eq} \\
\mtwo &=& \frac{\alpha_2}{\alphas}\mgluino\simeq 0.3\,\mgluino.
\label{mtwomgluino:eq}
\eea
These relations, or at least the first of them,
are commonly assumed in most
studies of the MSSM, even though they are not
necessary in the context of the model.

Another
relation which stems from minimal SUGRA and which is commonly
assumed is the equity of all the (soft) mass parameters of all the sleptons,
squarks, and typically also
Higgs bosons, to some common (scalar) mass parameter $\mzero$
at the GUT scale.
Renormalization effects
cause the masses of
color-carrying sparticles to become, at the $\mz$ scale,
typically by a factor of a few
heavier than the ones of the states with electroweak interactions only.
Often one also imposes a very attractive mechanism
of radiative electroweak symmetry breaking (EWSB), which provides additional
constraint on the parameters of the model, in particular relates the
SUSY Higgs/higgsino mass parameter $\mu$ to the parameters
of the model which break SUSY. This fully constrained framework has
been called the constrained MSSM (CMSSM).\cite{kkrw1} In practice,
various groups have considered the MSSM
with a varying number of additional assumptions, starting from
adopting just Eq.~(\ref{monemtwo:eq})
to the CMSSM with additional constraints,
\eg, from nucleon decay which requires specifying the underlying GUT,
or string, model, the simplest $SU(5)$ and $SO(10)$
models being the most commonly studied. (A discussion of GUT physics
is beyond the scope of
this talk and I will only occasionally make references to expected
corrections to low-energy variables, like $\alphasmz$, from simplest
GUT-models. ) It is not always easy to
discern what assumptions are actually responsible for what results. I
will make an attempt to sort some of those things out.

\section{Gauge Coupling Unification}

Gauge coupling unification has been perhaps the strongest
guiding principle behind the idea of GUT's.
Ever since the beginning it was known that, in the
framework of the SM, the scale of GUT's was about $\mx\sim10^{15}\gev$.
Supersymmetry causes this scale to
go up somewhat  ($\mx\sim10^{16}\gev$).
With improving accuracy of the experimental
data, especially $\ssqthw$,
it was becoming increasingly clear that the prediction of SUSY
was fitting
the data better than the one of the SM. But it wasn't until LEP turned on,
that, with precise data available for SM parameters, a final blow was given
to the idea of GUT's in the framework of the SM. In contrast, an impressive
confirmation of SUSY unification was pointed
out.\cite{earlyunif}

This early
work is of largely historical importance now, since during the last five
years a significant progress on both the experimental and theoretical sides
has been made. Initial studies, which used just 1-loop renormalization group
equations (RGE's) and a single SUSY breaking scale $\msusy$ were expanded to
include all dominant effects
and subleading corrections at both 1- and 2-loop level.
The most important effects are due to:
1-loop mass threshold corrections (which
correspond to
the fact that in reality masses of individual (s)particles of the MSSM are
typically non-degenerate);
2-loop (pure) gauge
contribution; and model-dependent corrections from GUT-scale physics. Each
of them provides $\sim10\%$ correction to the predicted value of $\alphas$,
assuming the electromagnetic constant $\alpha$ and
$\ssqthw$ as input parameters. Also important is the (inverse)
dependence of $\ssqthw$ on
$\mtop$,\cite{lp1} because the predicted $\alphasmz$ sensitively
depends on the allowed range of
$\ssqthw$.
Other contributions are
sub-dominant but are normally also included in any decent analysis. They
include: 2-loop Yukawa coupling contribution and mass-threshold
effects, and
scheme dependence ($\MS$ \vs\ $\DR$).
Recent updated discussions\cite{lp:new,ms} provide more
information.

Two general conclusions can be drawn from all those
extensive studies.
Firstly, gauge coupling unification in SUSY seems to be quite robust
in the sense that no large effects coming from electroweak or
GUT-scale physics exist which would destroy the picture, provided one
does not allow more than two Higgs doublets normally required in
SUSY. (However, fourth generation of fermions
is not in conflict with gauge unification.\cite{gunionforth})

Secondly, if one restricts oneself to masses roughly
below 1\tev\ then
generally
$\alphasmz\gsim0.12$\cite{kkrw1,lp1,lp:new} (and $\gsim0.13$ for SUSY
masses below some 300\gev).
This is because $\alphasmz$ grows
with decreasing masses of SUSY (s)particles (and also $\mtop$).
A recent updated analysis\cite{lp:new} quotes
$\alphasmz = 0.129 \pm 0.008$.  The limits quoted above include the
estimated theoretical errors
due to mass-thresholds at the GUT scale and higher-dimensional
non-renormalizable operators (NRO's)
in the GUT scale Lagrangian.
The above prediction for  $\alphasmz$ has been considered  a
success and the strongest evidence
in favor of supersymmetric unification,
especially in light of the range of $\alphasmz=0.127\pm0.05$
claimed by LEP experiments.\cite{langacker}

But the experimental status of $\alphasmz$
still remains unclear. All low-energy measurements and
lattice calculations of $\alphas$, when
translated to the scale $\mz$, generally give much lower values, between
0.11 and 0.117, with comparable or smaller error bars. (See, \eg,
recent reviews by Langacker\cite{langacker} for more detail.)
The only indication from low energies
for larger $\alphasmz$ from $\tau$ decays\cite{pich} has also been
questioned.\cite{shifman}

Very recently (in fact after the conference)
an interesting and important
development took place. First, Shifman\cite{shifman}
very vigorously argued that
the internal consistency of QCD requires that $\alphas$ be close to 0.11.
He gave a number of important reasons. Here I will quote only one:
large $\alphas$ $\sim0.125$ would correspond to
$\Lambda_{\MS}\approx500\mev$ (in contrast to $\sim200\mev$ for
$\alphasmz\approx0.11$). Such a large value is apparently in conflict with
crucial features of QCD on which a variety of phenomena depend
sensitively.
Prompted by Shifman's argument,
Voloshin\cite{voloshin} re-analyzed
$\Upsilon$ sum rules
claiming the record accuracy achieved so far: $\alphasmz = 0.109\pm
0.001$. On the other side, it has been argued\cite{consoli} that the
systematic error usually quoted in the LEP number is grossly
underestimated, and that at present LEP experiments can only claim
$0.10\lsim\alphasmz\lsim0.15$.

Clearly, small $\alphasmz\approx0.11$ seems an increasingly viable
possibility, while significantly larger values are predicted by the
CMSSM. Additionally, it has recently been shown\cite{bagger} that,
in the framework of the CMSSM, sub-leading quadratic corrections, play a
significant role when $\mhalf\lsim150\gev$,
and generally push up $\alphasmz$ obtained in the leading-log
(step-function) approximation, quoted above, by another
$\sim0.01$. This effect is
amplified by the mass relations (\ref{monemtwo:eq})--(\ref{mtwomgluino:eq})
among the gauginos.

The question arises as to whether SUSY unification can
accommodate $\alphasmz\approx0.11$, and at what expense.
Several solutions to this problem can be immediately suggested.
One is to remain
in the context of the CMSSM but adopt
a heavy SUSY scenario with the SUSY mass spectra significantly
exceeding
1\tev. This scenario would not only put SUSY into both theoretical
and experimental oblivion,
but is also, for the most part, inconsistent
with
our expectations that the lightest supersymmetric particle (LSP)
should be neutral
and/or with the lower bound on the age
of the Universe of at least some 10 billion
years.\cite{kkrw1} (See Section~5.)
Another possibility is to invoke large enough
negative corrections due to GUT-scale physics.
The issue was re-analyzed recently\cite{lp:new} and it was found that,
under natural assumptions,
$\alphasmz>0.12$.
Although it may well happen that the GUT-scale and NRO corrections
are abnormally large, the guiding idea of grand unification
becomes much less attractive in this case, and the predictive power
is essentially lost.

A different way out has been
suggested\cite{ms} lately: abandon the
additional assumptions, like
Eqs.~(\ref{monemtwo:eq})--(\ref{mtwomgluino:eq}), completely.
In fact, simple arguments\cite{ms} show
that it is just the relation (\ref{mtwomgluino:eq}) that is mainly
responsible for pushing $\alphasmz$ up so much in the CMSSM. By
relaxing it one can easily accommodate $\alphasmz\approx0.11$
for wide mass ranges, except for
the mass of the gluino which must remain rather low,
$\mgluino\lsim200-300\gev$. (The exact upper bound
depending on what upper limit on
$\alphasmz$ on adopts and how large GUT-physics corrections one is
willing to accept.) Furthermore, the wino mass parameter $\mtwo$ must
be large, $\mtwo\gsim3\mgluino$,
at least a few hundred \gev, in contrast to what is commonly
assumed.
This approach casts doubt also on the
relation~(\ref{monemtwo:eq}), which has been universally assumed in
phenomenological studies of charginos and
neutralinos, and which results from the same assumption at the GUT
scale.
(No ``direct'' constraint on $\mone$ can be placed from
the above considerations because the bino, being neutral, does not
enter the RGE's.)
Also, requiring
that the lightest (bino-like) neutralino be lighter than the gluino
leads to $\mone\lsim0.3\mtwo$, thus violating
the
relation~(\ref{monemtwo:eq}).\cite{dennis} It will be very
important to test Eqs.~(\ref{monemtwo:eq})--(\ref{mtwomgluino:eq})
in future experiments.\cite{fmpt}

To summarize, the issue of gauge coupling unification is still far
from being closed. Progress is needed on the experimental side to
determine more precisely $\alphasmz$. The most
commonly assumed ``minimal'' CMSSM scenario seems to imply too
high $\alphasmz$, at least if SUSY masses are below roughly 1\tev\ and
if $\alphasmz$ is indeed close to 0.11.
In fact, at present no {{\em simple} theoretical
framework motivated by GUT's, or strings, can really produce such
small $\alphasmz$.
It is also worth noting that it still
remains unclear how to reconcile the string unification
scale $M_{\rm string}\sim {\rm few}\times10^{17}\gev$ with the
$\mgut\sim10^{16}\gev$ suggested by bottom-up gauge coupling unification.

\section{Yukawa Coupling Unification}

Another driving idea, and a success, of the early GUT's was
the fact that the running masses of the b-quark and the $\tau$ were
apparently meeting at
roughly the gauge unification scale $\mx$. With more precise data on $\mb$
and $\mtau$ becoming available, it was shown that the \btau\
unification was only consistent in the SUSY framework, but not in the
non-SUSY case.\cite{ramondmbot} It is also badly spoiled if the fourth
generation of fermions exists.\cite{gunionforth}
Furthermore, several groups\cite{bbop,cpw,lp2}
pointed out that, in the MSSM-like framework, strict \btau\ unification is
implies a very heavy top $\mt=(200\gev)\sin\beta$. More precisely,
one needs large ($\sim1$) Yukawa coupling of the top-quark to balance
the effect of gauge couplings in the RGE's for $\mb/\mtau$. Requiring
perturbativity of the Yukawa couplings and imposing experimental
constraints on $\mb$ and $\mtau$ leads to
a rather narrow range of $\mt$ and $\tanb$. Furthermore, if $\mt$ is
to be also unified with $\mb$ and $\mtau$, like in simple versions of
$SO(10)$,
then large $\tanb\sim 50-60$ is needed.\cite{}

Two comments should be made in this context.
Recently, both the CDF and D0 collaborations have reported
discovery of the top
quark
and quoted:
$\mt=176 \pm8\pm10\gev$ (CDF)\cite{cdf:top} and $\mt= 199\pm
20\pm22\gev$
(D0).\cite{dzero:top}
Such large $\mt$ may be interpreted as supporting \btau\
unification but, unfortunately, does not help in constraining $\tanb$.

On the other hand, if $\alphasmz$ is small ($\sim0.11$), then
the above strong relation between $\tanb$ and $\mt$ can be
significantly relaxed\cite{kkrw1,ms}
provided that strict unification condition
$\mb/\mtau=1$ at
the GUT scale
is reduced somewhat ($\sim10\%$).
GUT-scale uncertainties of this size are actually expected in
GUT's.\cite{lp2} At the end, to test the idea of \btau\ unification,
we will need to constrain $\tanb$ independently. This probably won't be
possible before the couplings of the Higgs bosons are
precisely measured. The region of large $\tanb$ is also somewhat
sensitive to SUSY spectra.

\section{Mass Unification}

It is natural to expect that, in the unification framework, also the
various mass parameters in the MSSM will emerge from a few common sources.
Thus it has become customary to assume the so-called common gaugino
and common scalar masses, $\mhalf$ and $\mzero$, respectively,
at the GUT scale, and to consider the CMSSM (see Section~2).
These two assumptions are
certainly not irrefutable but are at
least sensible (except for the fact that assuming $\mhalf$ seems to be
in conflict\cite{ms} with
$\alphasmz\approx0.11$ -- see Section~3). They
obviously correspond to the simplest choice and furthermore
result from the simplest minimal supergravity framework.
Needless to say, most phenomenological studies of SUSY rely on at least
one of them, at least for the sake of reducing the otherwise
huge number of unrelated SUSY mass parameters.
(One also assumes
that the trilinear soft SUSY-breaking terms are equal to $\azero$ at $\mx$,
although this assumption has actually almost no bearing here.)

One can next
derive complete mass spectra of all the Higgs and supersymmetric particles by
running their 1-loop RGEs between $\mx$ and $\mz$.
In the CMSSM
the spectra are parametrized in terms of just a few basic parameters
which can be
conveniently choosen as: $\mt$, $\tanb$, $\mhalf$,
$\mzero$, and $\azero$.
One also needs to employ the full 1-loop effective Higgs potential in order
to properly implement the conditions for EWSB.
Next one imposes on the resulting mass spectra
mass limits from current direct
experimental searches and CLEO data on \bsg.\cite{hewett}
Also, the LSP must be neutral and its relic density must satisfy
$\Omega_{LSP}h_0^2<1$ to be consistent with
limits on the age of the Universe of at least 10 billion
years.

The resulting parameter space of the CMSSM, consistent with all the
above constraints, is remarkably constrained and leads to several
important predictions. I will describe them briefly relying on a
comprehensive analysis of Kane, \etal\cite{kkrw1} Several other groups
have also studied various aspects of mass unification and
obtained similar results.\cite{roberts,others}

1. The lightest neutralino {\em comes out} as the only neutral LSP.
(In the MSSM one
commonly {\em assumes} it to be the LSP.)
Furthermore, it is almost always a nearly pure bino -- which
has been shown\cite{chiasdm}
to be the only sensible candidate for dark matter (DM)
with $\Omega_{\bino}h_0^2\sim1$.
However, in this constrained scenario event rates for both direct and
indirect detection of (galactic halo) neutralino DM are rather
small.\cite{dkkw} This is not good news for testing the CMSSM
even in the next generation of DM detectors.

2. The relic density of bino-like LSP's grows with $\mhalf$ and
$\mzero$. These can then be constrained from above
(except for some rare cases) by $\Omega_{LSP}h_0^2<1$.\cite{roberts,kkrw1}
As a
result SUSY mass spectra come out below about 1\tev\ without imposing
an ill-defined naturalness constraint. The fact that the cosmological
constraint coincides with the expected SUSY breaking scale is remarkable.

3. The resulting mass range of the (bino-like) neutralino lies between
some 20\gev\ and 200\gev. Also, important mass relations ensue:
$2\mchi\approx\mcharone\approx\mneuttwo\approx0.3\mgluino$.

4. The lightest Higgs boson $h$ has couplings closely resembling those
of the SM Higgs. It's mass is expected around 80--120\gev. Clearly,
LEP~II will have an excellent chance of covering (most of) this mass
range, critically depending however on the beam energy.
Other Higgs bosons are typically very heavy and their couplings to
gauge bosons are strongly suppressed.

5. Typical masses of squarks and sleptons in the CMSSM
are unfortunately also
rather large and lie typically above the reach of the Tevatron and
LEP~II, respectively.
The gluino is not
well-constrained and can either be found at the Tevatron
($\sim200\gev$ mass range) or only after the LHC runs for a few years
($\sim1-2\tev$ range).

6. Because typical SUSY and charged Higgs masses are rather large,
SUSY contributions to \bsg\ are typically small, in agreement\cite{jim} with
CLEO data.\cite{hewett} QCD uncertainties are still large\cite{hewett}
but in the future one can use this rare
process as a powerful constraint on the CMSSM and other SUSY models.

Due to space and time limitations I must leave unmentioned many other
interesting topics. They can be found in several recent
reviews\cite{langacker,topten} and source articles.

\section{Conclusions}
While the general framework of supersymmetric grand unifications seems
remarkably attractive, much remains to be done and
understood. Constrained and predictive
frameworks, like the commonly used CMSSM, don't run into conflict with
any experimental or theoretical constraints, except perhaps
for predicting somewhat
large $\alphasmz$. It will be possible but also challenging to test
the CMSSM in future accelerator experiments, rare processes, and dark
matter searches.
The issue of \btau\ unification won't be
experimentally tested before properties of at least the lightest Higgs
are well measured and $\tanb$ is determined.
Many predictions of the CMSSM and other SUSY models
remain to be verified by future
experiments. And once, sooner or later, SUSY is discovered, it will
probably take many years before we go beyond the supersymmetric SM.



\section{References}


\begin{thebibliography}{9}

\bibitem{earlyunif}
U.~Amaldi, \etal,
\PLB{260}{91}{443};
J.~Ellis, \etal,
\PLB{260}{91}{131};
P.~Langacker and M.-X.~Luo, \PRD{44}{91}{817};
F.~Anselmo, \etal,
{\it Nuovo Cim.}
{\bf104A} (1991) 1817, and
{\bf105A} (1992) 581.

\bibitem{minisugra}
For reviews, see, \eg, H.-P.~Nilles, \PRT{110}{84}{1};
H.E.~Haber and G.L.~Kane, \PRT{117}{85}{75}.

\bibitem{kkrw1}
G.~Kane, C.~Kolda, L.~Roszkowski, and J.~Wells,
\PRD{49}{94}{6173}.

\bibitem{lp1}
P. Langacker and N. Polonsky, \PRD{47}{93}{4028}.

\bibitem{lp:new}
P.~Langacker and N.~Polonsky,
hep-ph/9503214.

\bibitem{ms}
L.~Roszkowski and M.~Shifman,
hep-ph/9503358
{}.

\bibitem{gunionforth}
J.F.~Gunion, D.W.~McKay, and
H.~Pois, \PLB{334}{94}{339}.

\bibitem{langacker}
P.~Langacker,
hep-ph/9412361 and
hep-ph/9411247.

\bibitem{pich}
See, \eg, F.~Le~Diberder and A.~Pich, \PLB{286}{92}{147}.

\bibitem{shifman}
M.~Shifman,
hep-ph/9501222 ({\it Mod. Phys. Lett.}, to appear).

\bibitem{voloshin}
M.~Voloshin,
hep-ph/9502224
{}.

\bibitem{consoli}
M.~Consoli and F.~Ferroni,
hep-ph/9501371.

\bibitem{bagger}
J.~Bagger, \etal,
hep-ph/9501277; D.~Pierce, talk at this meeting.

\bibitem{dennis} L.R. thanks Dennis Silverman for this remark.

\bibitem{fmpt}
J.L. Feng, H. Murayama, M.E. Peskin, and X. Tata, hep-ph/9502260.

\bibitem{ramondmbot}
H.~Arason, \etal,
\PRD{47}{93}{232};
A.~Giveon, \etal,
\PLB{271}{91}{138};
P.H.~Frampton, \etal,
\PLB{277}{92}{130}.
%

\bibitem{bbop}
V.~Barger, \etal,
\PLB{314}{93}{351}.

%
\bibitem{cpw}
M.~Carena, \etal,
\NPB{406}{93}{59}.
%
\bibitem{lp2}
P.~Langacker and N.~Polonsky, \PRD{49}{94}{1454}.

\bibitem{cdf:top}
F.~Abe, \etal, CDF Collaboration,
hep-ex/9503002.

\bibitem{dzero:top}
S.~Abachi, \etal, D0 Collaboration.
hep-ex/9503003.

\bibitem{roberts}
R.G.~Roberts and L.~Roszkowski, \PLB{309}{93}{329}.

\bibitem{others} See, \eg,
V.~Barger, \etal,
\PRD{49}{94}{4908}; S.~Kelley, \etal, \NPB{398}{93}{3}
H.~Baer, \etal,
\PRD{50}{94}{2148}; R.~Arnowitt and P.~Nath, \PRD{49}{94}{1479}.

%
\bibitem{chiasdm} L.~Roszkowski, \PLB{262}{91}{59}.

\bibitem{dkkw}
E.~Diehl, G.~Kane, C.~Kolda, and J.~Wells, hep-ph/9502399.

\bibitem{jim}
J.~Wells, talk at this meeting.

\bibitem{hewett}
J.~Hewett, talk at this meeting; Y.~Kwon, talk at this meeting.

\bibitem{topten}
H.E.~Haber, hep-ph/9308209;
H.~Murayama, hep-ph/9410285.





\end{thebibliography}
\end{document}